\documentstyle{article}

\begin{document}

\title{{\bf An Algorithm for the Optimal Consistent Approximation to a Pairwise Comparisons Matrix by Orthogonal Projections } }
\author{Waldemar W. Koczkodaj\thanks{%
this research has been partially supported by the Provincial Government
through the Northern Ontario Heritage Fund Corporation} \\
Computer Science, Laurentian University \\
Sudbury, Ontario P3E 2C6, Canada \\
waldemar@ramsey.cs.laurentian.ca \and Marian Or\l owski\thanks{%
this project has been partially supported by the Australian Research Council
grant A49602509 } \\
School of Information Systems, Queensland University of Technology \\
Brisbane, Q 4001, Australia \\
Orlowski@fit.qut.edu.au}
\date{Sudbury, Canada and Brisbane, Australia, \today}
\maketitle

\begin{abstract}

The algorithm for finding the optimal consistent approximation of an
inconsistent pairwise comaprisons matrix is based on a logarithmic
transformation of a pairwise comparisons matrix into a vector space with the
Euclidean metric. Orthogonal basis is introduced in the vector space. The
orthogonal projection of the transformed matrix onto the space formed by the
images of consistent matrices is the required consistent approximation.
\end{abstract}

\section{Triad Inconsistency in Pairwise Comparisons}

Triad inconsistency was introduced in \cite{Kocz93} and generalized in \cite
{DuKo94}. Its convergency analysis was published in \cite{HolKocz96}. The
reader's familiarity with \cite{HolKocz96} is assumed due to space
limitations. Only the essential concepts of the pairwise comparison method
are recalled here.

The method of pairwise comparisons was introduced in embryonic form by
Fechner (see \cite{Fech1860}) and after considerable extension, made popular
by Thurstone (see \cite{Thur27}). It can be used as a powerful inference
tool and knowledge acquisition technique in knowledge-based systems and data
mining.

For the sake of our exposition we define an $N \times N$ pairwise comparison
matrix simply as a square matrix $M=[m_{ij}]$ such that $m_{ij}>0$ for every 
$i,j=1, \ldots ,n$. A pairwise comparison matrix $M$ is called {\em %
reciprocal} if $m_{ij} = \frac{1}{m_{ji}}$ for every $i,j=1, \ldots ,n$
(then automatically $m_{ii}=1$ for every $i=1, \ldots ,n $). Let


\begin{center}
$M=\left| 
\begin{array}{cccc}
1 & m_{12} & \cdots & m_{1n} \\ 
\frac{1}{m_{12}} & 1 & \cdots & m_{2n} \\ 
\vdots & \vdots & \vdots & \vdots \\ 
\frac{1}{m_{1n}} & \frac{1}{m_{2n}} & \cdots & 1
\end{array}
\right|$
\end{center}

\noindent where $m_{ij}$ expresses an expert's relative preference of
stimuli $s_i$ over $s_j$.

A pairwise comparison matrix $M$ is called consistent if $m_{ij} \cdot
m_{jk}=m_{ik}$ for every $i,j,k=1, \ldots ,n$. While every consistent matrix
is reciprocal, the converse is false in general. Consistent matrices
correspond to the ideal situation in which there are exact values $s_1,
\ldots , s_n$ for the stimuli. The quotients $m_{ij}=s_i/s_j$ then form a
consistent matrix. Conversely, the starting point of the pairwise
comparisons inference theory is Saaty's theorem (see \cite{Saaty77}) which
states that for every $N \times N$ consistent matrix $M=[m_{ij}]$ there
exist positive real numbers $s_1, \ldots s_n$ such that $m_{ij}=s_i/s_j$ for
every $i,j=1, \ldots , n$. The vector $s=[s_1, \ldots s_n]$ is unique up to
a multiplicative constant. The challenge to the pairwise comparisons method
comes from the lack of consistency of the pairwise comparisons matrices
which arise in practice (while as a rule, all the pairwise comparisons
matrices are reciprocal). Given an $N \times N$ matrix $M$ which is not
consistent, the theory attempts to provide a consistent $N \times N$ matrix $%
C$ which differs from matrix $M$ ``as little as possible''. Algorithms for
reducing the triad inconsistency in pairwise comparisons can be
significantly improved by orthogonal projections.

\section{The Definition of triad $L-$consistency}

Let us recall that the matrices in the original space consist of positive
elements. The problem of the best approximation of a given matrix $M=[m_{ij}]
$ by a consistent matrix is transformed into a similar problem of
approximating a matrix $M^{\prime }=[\log m_{ij}]$ by a logarithmic image of
a consistent matrix. The benefit of such an approach is that the
logarithmically transformed images of consistent matrices form a linear
subspace {\bf $L$ } in $R^{N\times N}$. Each matrix in the subspace {\bf $L$ 
} is called a triad L-consistent matrix $M^{\prime }=[m_{ij}^{\prime }]$ and
satisfies the condition: $m_{ik}^{\prime }+m_{kj}^{\prime }=m_{ij}^{\prime }$
for every $i,j,k=1\ldots n$. It is much easier to work with linear spaces
and to use the tools of linear algebra than to work in manifolds
(topological or differential). Also the notion of closeness of matrices is
translated from one space to the other since the logarithmic transformation
is homeomorphic (one-to-one continuous mapping with a continuous inverse;
see \cite{Fundam}, Vol. II, page 593 for details). In other words two
matrices are close to each other in the sense of the Euclidean metric if
their logarithmic images are also close in the Euclidean metric.

Let us recall that matrices in the original space have all positive
elements. The approximation problem is reduced to the problem of finding the
orthogonal projection of the matrix $M^{\prime}$ on {\bf $L$ } since we opt
for the least square approximation in the space of logarithmic images of
matrices. The following algorithm is proposed to solve the above problem:

\begin{enumerate}
\item  Find a basis in {\bf $L$ }.

\item  Orthogonalize it (or orthonormalize it)

\item  Compute a projection $M^{\prime \prime }$ of $M^{\prime }$ on {\bf $L$
} using the orthonormal basis of {\bf $L$ } found in step 2.
\end{enumerate}


Steps (1) and (2) produce a basis of the space {\bf $L$ } . This is done
once only for a matrix of a given size $N$ . A Gram-Schmidt
orthogonalization procedure is used for constructing an orthogonal basis in 
{\bf $L$ } . The actual algorithm for finding a triad $L-$consistent
approximation is based on step (3); therefore the approximation problem is
reduced to:

\noindent {\em given a matrix A (a logarithmic image of the matrix to be
approximated by a consistent matrix), find the orthogonal projection $%
A^{\prime}$ of $A$ onto {\bf $L$ }.}

The most natural way of solving this problem is to project the matrix $A$ on
the one dimensional subspaces of {\bf $L$ } generated by each vector in the
orthogonal basis of {\bf $L$ } and then sum these projections. While most of
the computation is routine, the problem of finding an orthogonal basis in
the space {\bf $L$ } is somewhat challenging. For every $N\times N$
consistent matrix $A$ there exists a vector of stimuli $(s_{1},s_{2},\dots
,s_{N})$, unique up to a multiplicative constant such that $a_{ij}=\frac{%
s_{i}}{s_{j}}$. One may thus infer that the dimension of the space {\bf $L$ }
is $N-1$. As a consequence, this observation stipulates that the space {\bf $%
L$ } has to have a basis comprised of $N-1$ elements.

Analysis of numerous examples has led to the discovery of the following
basis matrices $B_k = [b^k_{ij}]$

\[
b^k_{ij} = \left\{ 
\begin{array}{r@{\quad}l}
1, & $for $1 \le i \le k < j \le N \\ 
-1, & $for $1 \le j \le k < i \le N \\ 
0, & $otherwise $%
\end{array}
\right. 
\]

Fig. 1 illustrates the basis matrices for $N=7$. In essence each basis
matrix $B_{k}$ contains two square blocks of $0s$ (situated symmetrically
about the main diagonal) of size $k$ and $N-k$, a block of $1s$ of size $k$
by $N-k$ above the main diagonal, and a block of $-1s$ of size $N-k$ by $k$
below the main diagonal, where $k=1,\ldots ,N-1$. \\[1ex]

\setlength{\tabcolsep}{4pt} 
\begin{tabular}{r|rp{8pt}p{8pt}p{8pt}rrr|p{3pt}r|rrp{8pt}p{8pt}p{8pt}rr|p{3pt}r|rrrp{8pt}p{8pt}p{8pt}r|}
& 0 & 1 & 1 & 1 & 1 & 1 & 1 &  &  & 0 & 0 & 1 & 1 & 1 & 1 & 1 &  &  & 0 & 0
& 0 & 1 & 1 & 1 & 1 \\ 
& -1 & 0 & 0 & 0 & 0 & 0 & 0 &  &  & 0 & 0 & 1 & 1 & 1 & 1 & 1 &  &  & 0 & 0
& 0 & 1 & 1 & 1 & 1 \\ 
& -1 & 0 & 0 & 0 & 0 & 0 & 0 &  &  & -1 & -1 & 0 & 0 & 0 & 0 & 0 &  &  & 0 & 
0 & 0 & 1 & 1 & 1 & 1 \\ 
$B_1$= & -1 & 0 & 0 & 0 & 0 & 0 & 0 &  & $B_2$= & -1 & -1 & 0 & 0 & 0 & 0 & 0
&  & $B_3$= & -1 & -1 & -1 & 0 & 0 & 0 & 0 \\ 
& -1 & 0 & 0 & 0 & 0 & 0 & 0 &  &  & -1 & -1 & 0 & 0 & 0 & 0 & 0 &  &  & -1
& -1 & -1 & 0 & 0 & 0 & 0 \\ 
& -1 & 0 & 0 & 0 & 0 & 0 & 0 &  &  & -1 & -1 & 0 & 0 & 0 & 0 & 0 &  &  & -1
& -1 & -1 & 0 & 0 & 0 & 0 \\ 
& -1 & 0 & 0 & 0 & 0 & 0 & 0 &  &  & -1 & -1 & 0 & 0 & 0 & 0 & 0 &  &  & -1
& -1 & -1 & 0 & 0 & 0 & 0
\end{tabular}
\\[1ex]

\begin{tabular}{r|rrrrrrr|p{3pt}r|rrrrrrr|p{2pt}r|rrrrrrr|}
& 0 & 0 & 0 & 0 & 1 & 1 & 1 &  &  & 0 & 0 & 0 & 0 & 0 & 1 & 1 &  &  & 0 & 0
& 0 & 0 & 0 & 0 & 1 \\ 
& 0 & 0 & 0 & 0 & 1 & 1 & 1 &  &  & 0 & 0 & 0 & 0 & 0 & 1 & 1 &  &  & 0 & 0
& 0 & 0 & 0 & 0 & 1 \\ 
& 0 & 0 & 0 & 0 & 1 & 1 & 1 &  &  & 0 & 0 & 0 & 0 & 0 & 1 & 1 &  &  & 0 & 0
& 0 & 0 & 0 & 0 & 1 \\ 
$B_4$= & 0 & 0 & 0 & 0 & 1 & 1 & 1 &  & $B_5$= & 0 & 0 & 0 & 0 & 0 & 1 & 1 & 
& $B_6$= & 0 & 0 & 0 & 0 & 0 & 0 & 1 \\ 
& -1 & -1 & -1 & -1 & 0 & 0 & 0 &  &  & 0 & 0 & 0 & 0 & 0 & 1 & 1 &  &  & 0
& 0 & 0 & 0 & 0 & 0 & 1 \\ 
& -1 & -1 & -1 & -1 & 0 & 0 & 0 &  &  & -1 & -1 & -1 & -1 & -1 & 0 & 0 &  & 
& 0 & 0 & 0 & 0 & 0 & 0 & 1 \\ 
& -1 & -1 & -1 & -1 & 0 & 0 & 0 &  &  & -1 & -1 & -1 & -1 & -1 & 0 & 0 &  & 
& -1 & -1 & -1 & -1 & -1 & -1 & 0
\end{tabular}
\\[1ex]

\begin{center}
Fig 1. An example of a basis of {\bf $L$ } for $N = 7$
\end{center}

\vspace{2ex}

\noindent {\bf Proposition 1.} The matrices $B_k$ are linearly independent. 
\\[0.5ex]

\noindent {\bf Proof.} The rank of the following matrix containing as its
rows the $enlisted$ (by rows) matrices $B_k$ is $N-1$ because the
determinant of the submatrix formed by column $12, 13, \ldots , 1N$ is equal
to 1: 

\begin{tabular}{cccccccccccccccc}
&  &  &  &  &  &  &  &  &  &  &  &  &  &  &  \\ 
& 11 & 12 & 13 & 14 & \ldots & 1N & 21 & 22 & 23 & \ldots & N1 & N2 & N3 & 
\ldots & NN \\ \hline
&  &  &  &  &  &  &  &  &  &  &  &  &  &  &  \\ 
$B_1$ & 0 & 1 & 1 & 1 & \ldots & 1 & -1 & 0 & 0 & \ldots & -1 & 0 & 0 & 
\ldots & 0 \\ 
$B_2$ & 0 & 0 & 1 & 1 & \ldots & 1 & 0 & 0 & 1 & \ldots & -1 & -1 & 0 & 
\ldots & 0 \\ 
\ldots & \ldots & \ldots & \ldots & \ldots & \ldots & \ldots & \ldots & 
\ldots & \ldots & \ldots & \ldots & \ldots & \ldots & \ldots & \ldots \\ 
$B_{N-1}$ & 0 & 0 & 0 & 0 & \ldots & 1 & 0 & 0 & 0 & \ldots & -1 & -1 & -1 & 
\ldots & 0
\end{tabular}
\\[2ex]

\noindent {\bf Proposition 2.} In an antisymmetric matrix, the set of
conditions:\newline

\begin{enumerate}
\item  $x_{pq}+x_{qs}=x_{ps}$ where $p,q,s$ are pairwise different \noindent
is equivalent to:\newline

\item  $x_{ij}+x_{jk}=x_{ik}$ where $i<j<k$.
\end{enumerate}

\vspace{1ex}

\noindent {\bf Proof.} Let us assume that $s$ is between $p$ and $q$. If $p
< q$ then $(1)$ can be written as: \newline
\[
x_{pq} \: = \: x_{ps} \: - \: x_{qs} 
\]

\noindent and by symmetry: \newline
\[
x_{pq} \: = \: x_{ps} \: + \: x_{qs} 
\]

\noindent which is exactly $(2)$ if we set $(p,s,q) = (i,j,k)$. \\[0.5ex]

The reasoning in other cases (for $p$ or $q$ in the middle) is the same
because of the symmetry of condition $(1)$ with respect to the coefficients $%
(p,q,s)$.

\vspace{2ex}

As a consequence of Proposition 2 we do not need to check all matrix
elements. It is enough to check the elements above the main diagonal.
Proposition 2 is used in the proof of Proposition 3.

Let us now check if the proposed basis matrices are triad $L-$consistent. It
is sufficient (in light of the above proposition and because of their
symmetry) to check that they are triad $L-$consistent with respect to the
entries above the diagonal.

\vspace{2ex}

\noindent {\bf Proposition 3.} All matrices $B_k$ satisfy the following
condition: 
\[
x_{ij} + x_{jk} = x_{ik} \quad {\em where} \:i < j < k 
\]

\noindent {\bf Proof.} The above condition stipulates that the value in the
right upper corner of the 
rectangle (see Fig. 2) is the sum of values from the left-upper and bottom
right corner:

\begin{center}
\setlength{\unitlength}{1.0cm} 
\begin{picture}(6.5,6.5)
\put(2,6.2){\bf $i$}          
\put(3,6.2){\bf $j$}          
\put(5,6.2){$k$}              
\put(0,4){\bf $i$}            
\put(0,3){\bf $j$}            
\put(0,1){$k$}                

\put(4.5,4){\bf $1$}          
\put(4.5,3){\bf $1$}          

\put(1,5){\bf $0$}
\put(0.5,5.5){\bf $0$}
\put(1.75,4.25){\bf $\cdot$}
\put(1.25,4.75){\bf $\cdot$}
\put(1.5,4.5){\bf $\cdot$}

\put(2,4){\bf $0$}
\put(3,4){\bf $\ast$}
\put(5,4){\bf $\ast$}
\put(3,3){\bf $0$}
\put(4,2){\bf $0$}
\put(3.75,2.25){\bf $\cdot$}
\put(3.5,2.5){\bf $\cdot$}

\put(5,3){\bf $\ast$}
\put(5,1){\bf $0$}
\put(5.5,0.5){\bf $0$}

\put(4.5,2){\bf $1$}   
\put(5,2){\bf $1$}     
\put(5.5,2){\bf $1$}   

\put(4.5,1.5){\bf $0$}

\put(4.5,5.5){\bf $1$} 
\put(5,5.5){\bf $1$}
\put(5.5,5.5){\bf $1$}
\put(4.5,5){\bf $1$} 
\put(5,5){\bf $1$}
\put(5.5,5){\bf $1$}

\put(5.5,4){\bf $1$} 
\put(5.5,3){\bf $1$} 

\put(0.3,0.4)  {\framebox(5.5,5.6)}     
\put(2.75,2.85){\framebox(2.6,1.55)}    
\put(4.4,1.9)  {\framebox(1.4,4.1)}     
\put(0.3,4.1){\line(1,0){1.6}}          
\put(2.1,4.4){\line(0,1){1.6}}          
\put(2.25,4.1){\line(1,0){0.5}}         
\put(3.3,4.1){\line(1,0){1.06}}         
\put(4.7,4.1){\line(1,0){0.25}}         
\put(5.35,4.1){\line(1,0){0.15}}        

\put(0.3,1.1){\line(1,0){4.6}}          

\put(5.1,5.85){\line(0,1){.15}}          
\put(5.1,5.3){\line(0,1){.1}}          
\put(5.1,4.4){\line(0,1){.5}}            
\put(5.1,3.3){\line(0,1){.6}}            
\put(5.1,2.85){\line(0,-1){.5}}          
\put(5.1,1.9){\line(0,-1){.5}}           

\put(0.3,3.1){\line(1,0){2.45}}          
\put(3.1,4.4){\line(0,1){1.6}}           
\put(3.3,3.1){\line(1,0){1.06}}          
\put(4.7,3.1){\line(1,0){0.25}}          
\put(5.35,3.1){\line(1,0){0.15}}         

\end{picture}

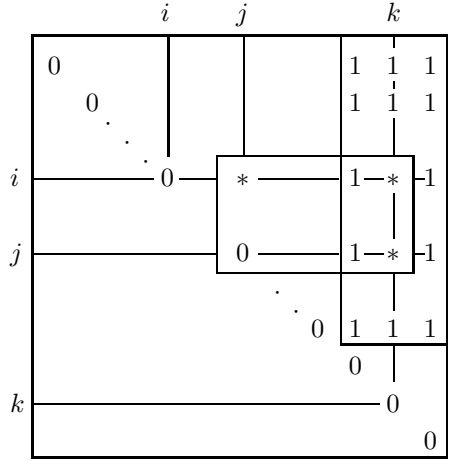
Fig 2. Partitioning the matrix \\[3ex]
\end{center}

\noindent Each of the basis matrices satisfies this condition. There are two
cases to be considered:

\begin{itemize}
\item  case 1 - the ``starred'' corners are outside the rectangle of $1s$ in
the matrix $B_{k}$.

\item  case 2 - a ``starred'' corner is in the rectangle of $1s$ in the
matrix $B_{k}$.
\end{itemize}

\noindent In case 1, the entries in ``starred'' corners are all $0s$ and the
condition in question is satisfied since: 
\[
x_{ij} + x_{jk} = 0 + 0 = 0 = x_{ik} 
\]

\noindent In Case 2, always two (but never three) ``starred'' corners are in
the rectangle of $1s$ of $B_k$ and one of them is $x_{ik}$. Therefore the $%
LHS$ of the expression is $1$ and so is the $RHS$. 
\[
x_{ij} + x_{jk} = 0 + 1 = 1 = x_{ik} 
\]
\noindent or by symmetry 
\[
x_{ij} + x_{jk} = 1 + 0 = 1 = x_{ik} 
\]

\vspace{2ex}

\noindent {\bf Proposition 4.}. A linear combination of triad $L-$consistent
matrices is triad $L-$consistent. \\[1ex]

\noindent {\bf Proof.} This follows from elementary algebra. A linear
combination of objects satisfying a linear condition in Proposition 3
satisfy the same condition, i.e.:

\noindent if $x_{ij}+x_{jk}=x_{ik}$ and $y_{ij}+y_{jk}=y_{ik}$ then $%
(ax_{ij}+by_{ij})+(ax_{jk}+by_{jk})=ax_{ik}+by_{ik}$

\vspace{2ex}

\noindent The above considerations lead to formulation of the following
Theorem. \\[0.5ex]

\noindent {\bf Theorem.} Every triad $L-$consistent $N \times N$ matrix is a
linear combination of the basis matrices $B_k = [b^k_{ij}]$ for $k = 1,2,
\dots ,N-1$.

\vspace{2ex}

\section{The Orthogonalization Algorithm}

The Gram-Schmidt orthogonalization process (see, for example, \cite{Fundam})
can be used to construct the basis. 
The fairly ``regular'' form of the basis matrices suggests that the
orthogonal basis should also be quite regular. Indeed, solving a system of $%
N-2$ linear equations produces the following orthogonal basis matrices $T_k
= [t^k_{ij}]$:

\[
t^k_{ij} = \left\{ 
\begin{array}{r@{\quad}l}
-\frac{N-k}{N-k+1} & $for \ $i < k = j \\ 
\frac{1}{N-k+1} & $if \ $ i < k < j \le N \\ 
1 & $if \ $ i = k < j \le N \\ 
-t^k_{ji} & $if \ $ t_{ij} \neq 0 \ and \ j < i \\ 
0 & $otherwise $%
\end{array}
\right. 
\]
\\[0.5ex]

\noindent This is equivalent to the following simpler non-recursive
definition:

\[
T_k = B_k - \frac{N-k}{N-k+1} B_{k-1} 
\]

\noindent where $B_0$ is a matrix with all zero elements. \\[1ex]

Space limitations force the authors to rely on the reader's knowledge of
basic linear algebra to limit this presentation to the final formula and an
example. However, the detailed computation leading to the above formula is
available by internet in each author's WEB page. The orthogonal basis for
the case $N = 7$ is presented in Fig. 3.\\[1ex]

\setlength{\arraycolsep}{4pt} 
\[
T_1= \left| 
\begin{array}{rrrrrrr}
0 & 1 & 1 & 1 & 1 & 1 & 1 \\ 
-1 & 0 & 0 & 0 & 0 & 0 & 0 \\ 
-1 & 0 & 0 & 0 & 0 & 0 & 0 \\ 
-1 & 0 & 0 & 0 & 0 & 0 & 0 \\ 
-1 & 0 & 0 & 0 & 0 & 0 & 0 \\ 
-1 & 0 & 0 & 0 & 0 & 0 & 0 \\ 
-1 & 0 & 0 & 0 & 0 & 0 & 0
\end{array}
\right| \:\:T_2= \left| 
\begin{array}{rrrrrrr}
0 & -\frac{5}{6} & \frac{1}{6} & \frac{1}{6} & \frac{1}{6} & \frac{1}{6} & 
\frac{1}{6} \\ 
\frac{5}{6} & 0 & 1 & 1 & 1 & 1 & 1 \\ 
-\frac{1}{6} & -1 & 0 & 0 & 0 & 0 & 0 \\ 
-\frac{1}{6} & -1 & 0 & 0 & 0 & 0 & 0 \\ 
-\frac{1}{6} & -1 & 0 & 0 & 0 & 0 & 0 \\ 
-\frac{1}{6} & -1 & 0 & 0 & 0 & 0 & 0 \\ 
-\frac{1}{6} & -1 & 0 & 0 & 0 & 0 & 0
\end{array}
\right| \:\:T_3= \left| 
\begin{array}{rrrrrrr}
0 & 0 & -\frac{4}{5} & \frac{1}{5} & \frac{1}{5} & \frac{1}{5} & \frac{1}{5}
\\ 
0 & 0 & -\frac{4}{5} & \frac{1}{5} & \frac{1}{5} & \frac{1}{5} & \frac{1}{5}
\\ 
\frac{4}{5} & \frac{4}{5} & 0 & 1 & 1 & 1 & 1 \\ 
-\frac{1}{5} & -\frac{1}{5} & -1 & 0 & 0 & 0 & 0 \\ 
-\frac{1}{5} & -\frac{1}{5} & -1 & 0 & 0 & 0 & 0 \\ 
-\frac{1}{5} & -\frac{1}{5} & -1 & 0 & 0 & 0 & 0 \\ 
-\frac{1}{5} & -\frac{1}{5} & -1 & 0 & 0 & 0 & 0
\end{array}
\right| 
\]

\[
T_4= \left| 
\begin{array}{rrrrrrr}
0 & 0 & 0 & -\frac{3}{4} & \frac{1}{4} & \frac{1}{4} & \frac{1}{4} \\ 
0 & 0 & 0 & -\frac{3}{4} & \frac{1}{4} & \frac{1}{4} & \frac{1}{4} \\ 
0 & 0 & 0 & -\frac{3}{4} & \frac{1}{4} & \frac{1}{4} & \frac{1}{4} \\ 
\frac{3}{4} & \frac{3}{4} & \frac{3}{4} & 0 & 1 & 1 & 1 \\ 
-\frac{1}{4} & -\frac{1}{4} & -\frac{1}{4} & -1 & 0 & 0 & 0 \\ 
-\frac{1}{4} & -\frac{1}{4} & -\frac{1}{4} & -1 & 0 & 0 & 0 \\ 
-\frac{1}{4} & -\frac{1}{4} & -\frac{1}{4} & -1 & 0 & 0 & 0
\end{array}
\right| \:\:T_5= \left| 
\begin{array}{rrrrrrr}
0 & 0 & 0 & 0 & -\frac{2}{3} & \frac{1}{3} & \frac{1}{3} \\ 
0 & 0 & 0 & 0 & -\frac{2}{3} & \frac{1}{3} & \frac{1}{3} \\ 
0 & 0 & 0 & 0 & -\frac{2}{3} & \frac{1}{3} & \frac{1}{3} \\ 
0 & 0 & 0 & 0 & -\frac{2}{3} & \frac{1}{3} & \frac{1}{3} \\ 
\frac{2}{3} & \frac{2}{3} & \frac{2}{3} & \frac{2}{3} & 0 & 1 & 1 \\ 
-\frac{1}{3} & -\frac{1}{3} & -\frac{1}{3} & -\frac{1}{3} & -1 & 0 & 0 \\ 
-\frac{1}{3} & -\frac{1}{3} & -\frac{1}{3} & -\frac{1}{3} & -1 & 0 & 0
\end{array}
\right| \:\:T_6= \left| 
\begin{array}{rrrrrrr}
0 & 0 & 0 & 0 & 0 & -\frac{1}{2} & \frac{1}{2} \\ 
0 & 0 & 0 & 0 & 0 & -\frac{1}{2} & \frac{1}{2} \\ 
0 & 0 & 0 & 0 & 0 & -\frac{1}{2} & \frac{1}{2} \\ 
0 & 0 & 0 & 0 & 0 & -\frac{1}{2} & \frac{1}{2} \\ 
0 & 0 & 0 & 0 & 0 & -\frac{1}{2} & \frac{1}{2} \\ 
\frac{1}{2} & \frac{1}{2} & \frac{1}{2} & \frac{1}{2} & \frac{1}{2} & 0 & 1
\\ 
-\frac{1}{2} & -\frac{1}{2} & -\frac{1}{2} & -\frac{1}{2} & -\frac{1}{2} & -1
& 0
\end{array}
\right| 
\]

\begin{center}
Fig. 3. An example of orthogonal basis for $N = 7$
\end{center}

\vspace{2ex}

\noindent The Euclidean norms of the basis matrices can be computed by the
following formula:

\[
|T_k|^2 = 2\{(k-1)*\left[\frac{N-k}{(N-k+1)^2} + \frac{(N-k)^2}{(N-k+1)^2}
\right] + N-k\} = \frac{2(N-k)}{(N-k+1)} 
\]

The orthogonal basis for the space {\bf $L$ } is given above (see the
formulas for $t_{ij}$ and $T_k$) and for a given $N$ one can produce the $%
N-1 $ matrices $T_k$. Once the matrices $T_k$ are determined we may compute
the following values for a given matrix $A$ (note that operation $\cdot$ is
a dot product; not a regular matrix product): 
\[
\forall (k = 1, \dots ,N-1): t_k = \frac{T_k \cdot A}{|T_k|^2} 
\]

\noindent The next step is to compute the linear combination 
\[
A^{\prime}= \sum_{k=1}^{N-1} t_k \times T_k 
\]

\noindent where the operation $\times$ is a scalar multiplication.

\vspace{2ex}

The result is the required projection of A into {\bf $L$ }. It is easy to
see that the complexity of computing the coefficients $t_k$ and hence the
matrix $A^{\prime}$ is $O(n^2)$.

\section{Conclusions}

The triad inconsistency definition provides an opportunity for reducing the
inconsistency of the experts' judgements. It can also be used as a technique
for data validation in the knowledge acquisition process. The inconsistency
measure of a comparison matrix can serve as a measure of the validity of the
knowledge.

The technique presented here for calculating a consistent approximation to a
pairwise comparisons matrix is an important step forward. The use of an
orthogonal basis simplifies the computation of the mapping of a given matrix
since it is just a linear combination of the basis matrices which need be
computed only once for a problem of a given size.

A convincing argument for using an orthogonal basis is a consideration of
the complication that arises in ordinary geometry when oblique axes are used
instead of orthogonal axes.

The use of an orthogonal basis leads to an algorithm that is simple to
implement (especially in a language supporting matrix operations).

\end{document}